\documentclass[twocolumn, twocolappendix]{aastex701}

\usepackage{amsmath}
\usepackage{booktabs}
\newcommand{\mr}{\mathrm}
\usepackage{physics}
\usepackage{amssymb}

\newcommand{\rp}{{\textit{r}-process}}

\begin{document}

\title{Binary Neutron Star Merger Evolution and \textit{r}-Process Enrichment in the Milky Way Disk}

\author[orcid=0009-0008-4809-3775, gname=Joon Young,sname=Lee]{Joon Young Lee}
\affiliation{Department of Physics, The University of Texas at Austin, 2515 Speedway, Austin, TX 78712, USA}
\email{jylee0902@utexas.edu}  

\author[orcid=0000-0001-5403-3762, gname=Hsin-Yu,sname=Chen]{Hsin-Yu Chen} 
\affiliation{Department of Physics, The University of Texas at Austin, 2515 Speedway, Austin, TX 78712, USA}
\email{hsinyu@austin.utexas.edu}

\author[orcid=0000-0002-3836-7751, gname=Muhammed,sname=Saleem]{Muhammed Saleem}
\affiliation{Department of Physics, The University of Texas at Austin, 2515 Speedway, Austin, TX 78712, USA}
\email{muhammed.cholayil@austin.utexas.edu}

\author[orcid=0000-0001-6374-6465, gname=Daniel M.,sname=Siegel]{Daniel M. Siegel}
\affiliation{Institute of Physics, University of Greifswald, D-17489 Greifswald, Germany}
\email{daniel.siegel@uni-greifswald.de}

\begin{abstract}
\noindent

The origin of half of the rapid neutron-capture (\textit{r}-process) nucleosynthesis elements in the Universe remains an open question. Binary neutron star (BNS) mergers have been shown to face difficulties in reproducing the observed \textit{r}-process enrichment in Milky Way disk stars. However, their \textit{r}-process enrichment efficiency may evolve with redshift beyond the star formation history, potentially due to evolution in the merger rate or the average \textit{r}-process yield in the early Universe. Such evolution could help reconcile BNS mergers with the observed abundance patterns in disk stars. In this paper, we explore the possibility that BNS mergers with an evolving enrichment efficiency could serve as the sole \textit{r}-process production channel. By jointly comparing gravitational-wave observations from LIGO--Virgo--KAGRA (LVK), short gamma-ray bursts, Galactic neutron stars, and stellar abundance measurements in Milky Way disk stars, we find that scenarios with additional evolution are strongly preferred over non-evolving scenarios to reproduce the abundance patterns, with Bayes factors exceeding $10^{20}$. We quantify the required evolution in both the merger rate and yield, and directly compare them with observations and theoretical predictions. We find that the evolving scenarios are in tension with short gamma-ray burst observations and predictions from multiple population synthesis models, while remaining consistent with current stochastic gravitational-wave background constraints. Our results provide a quantitative framework for evaluating whether BNS mergers with evolving enrichment efficiency can account for the observed \textit{r}-process enrichment history of Milky Way disk stars.

\end{abstract}

\section{Introduction} 

    The origin of the heavy elements produced by rapid neutron capture (\textit{r}-process) remains a central problem in nuclear astrophysics. Compact-object mergers, especially binary neutron star (BNS) mergers, are among the most promising candidate sites because they provide neutron-rich ejecta and have direct observational support from GW170817 and its associated kilonova \citep{Freiburghaus_1999, Metzger_2010, RJ_Foley, Siegel_2019_2,vieira2025spectroscopicrprocessabundanceretrieval}. 

    Valid \rp\ production channels must be able to reproduce the observed amount of \rp\ material in the Universe and their evolution across the history. The observations of stellar chemical abundances are good tracers of the \rp\ enrichment history (see e.g., \citealt{Burris_2000, Cescutti_2006, Battistini_2016, Wehmeyer_2015,Hotokezaka_2018}). Several studies using stellar abundance observations have concluded that BNS mergers alone are insufficient to explain the \rp\ element abundances observed in the Milky Way \citep{Haynes_2018, Hotokezaka_2018,Siegel_2019_2, Siegel_2019_1, Cote_2019, molero2025,Chen_2025}, attributing this tension to the long delay times of BNS mergers relative to the star formation history (SFH).
    
    Recently, \citet{Chen_2025}, and a follow-up study by \citet{saleem2026}, developed a multi-messenger Bayesian framework that combines gravitational-wave (GW) constraints on BNS merger rates, pulsar observations, delay time distributions (DTDs) from short gamma-ray bursts (sGRBs), and neutron star equation of state (EOS) constraints within a one-zone Galactic chemical evolution model. This approach enables a self-consistent treatment of uncertainties in both observations and models, including merger rates and \textit{r}-process yields. Their analysis showed that, although the inferred local BNS rate and yield are broadly consistent with existing astrophysical and geophysical constraints, a BNS-only scenario with a sGRB-informed DTD fails to reproduce the declining $[\mr{Eu/Fe}]$\footnote{$\mr{[A/B]} = \mr{log}_{10}(N_\mr{A}/N_\mr{B})-\mr{log}_{10}(N_\mr{A}/N_\mr{B})_\odot$ denotes the logarithmic abundance ratios relative to the Sun.} trend observed in Milky Way disk stars. Instead, either an unrealistically short delay time is required, which disagrees with the standard binary evolution models \citep{Dominik_2012, Giacobbo_2018, 2018MNRAS.481.4009V} and sGRB observations \citep{Zevin_2022,Behroozi_2014}—or an additional \textit{r}-process source that more closely tracks the SFH must contribute significantly. To explore this possibility, \citet{saleem2026} tested merger-based alternatives, including fast-merging BNS systems and neutron star--black hole (NSBH) mergers, and found that neither can account for the observed abundance evolution without invoking extreme and implausible assumptions, thereby reinforcing the need for an additional, non-merger-dominated enrichment channel.

    The aforementioned studies assumed that the BNS merger rate follows the SFH convolved with a fixed DTD. However, it remains an open question whether the BNS merger rate may have evolved differently due to effects such as a metallicity-dependent BNS formation efficiency. Another possibility would be the evolution of the \textit{r}-process yields, which can be caused by the change of the mass and spin distribution of NSs along cosmic time. Several studies \citep{van_Son_2025, Broekgaarden_2022, Santoliquido_2021, Neijssel_2019} have shown that compact binaries in general, and binary black holes (BBH) and NSBH binaries in particular, are formed more efficiently at lower metallicities. This implies that, despite the delay time, stars formed before the peak of cosmic star formation ($z\sim 2$) may be more likely to produce compact binaries than those formed after the peak. Although these studies find no clear evidence for an evolving formation efficiency in BNSs, their conclusions remain subject to substantial uncertainties, and the possibility of such evolution cannot be ruled out. This is the main context for the present work. 
    
    In this work, building upon the previous works by \citet{Chen_2025} and \citet{saleem2026} we investigate the possibility of BNS \rp\ enrichment efficiency evolving with redshift and beyond SFH. We parameterize this evolution as a broken power-law in $(1+z)$, as a phenomenological choice to allow the efficiency to vary differently at different redshifts. This choice provides a description of any effective redshift dependence in the contribution of BNS mergers to Galactic \textit{r}-process enrichment. In other words, the modeled evolution can manifest various underlying causes such as metallicity-dependent formation history or a redshift dependence in the average yield itself. The framework therefore fully explores whether the effective evolution of the BNS enrichment can describe the Milky Way's observed chemical evolution in the disk.
    
    The rest of this paper is organized as follows.
    In Section~\ref{sec:method}, we describe the Galactic chemical evolution model and the Bayesian inference framework.
    In Section~\ref{sec:results}, we present the results of our scenarios and compare them with observations and theoretical expectations.
    Finally, in Section~\ref{sec:discussion}, we discuss the implications of our results, outline limitations, and conclude with prospects for future work.

\section{method}
\label{sec:method}
    We use a one-zone Galactic chemical evolution model which considers various enrichment channels as inputs and provides the stellar abundance evolution curves over cosmic history as the output~\citep{Siegel_2019_1}. This model captures the average level of chemical enrichment in the interstellar medium (ISM) at high metallicity, where the ISM has been enriched by a large number of individual events, as appropriate for the Milky Way disk stars considered in our study. The channels include \textit{r}-process producing channels such as BNS mergers, and iron enrichment channels such as CCSNe and Type Ia SNe. Each channel is primarily described by its rate and yield. 
    We provide more details of the framework below.

\subsection{Enrichment Sources}
    In this study, we follow the chemical evolution of two representative elements, Fe and Eu. Iron serves as a tracer of the overall metal enrichment driven primarily by supernova activity, while europium is adopted as a standard tracer for \textit{r}-process nucleosynthesis. Evolving these two elements together allows the reproduction of a abundance trend $\mr{([Fe/H],[Eu/Fe])}$, which not only constrains the amount of \rp\ production, but also the timing of the production with respect to metallicity. The production of iron is governed by CCSNe and SNe~Ia, assuming fixed local rate and fixed yield. This is an appropriate assumption, as their uncertainties are relatively small compared to those of other parameters, such as the rate and yield of BNS mergers (see more details in Appendix \ref{sub: iron production}).

    Diverse astrophysical sites have been proposed as possible sources of \textit{r}-process nucleosynthesis, including BNS and NSBH mergers \citep{1974ApJ...192L.145L}, collapsars \citep{Siegel_2019_1}, accretion-induced collapse of white dwarfs \citep{Wheeler_1998, Fryer_1999}, magnetar giant flares \citep{Patel_2025}, magnetorotational supernovae \citep{Winteler_2012, Yong_2021}, post-explosion NS wind \citep{Prasanna_2024}, and common envelope jets supernovae \citep{10.1093/mnras/stv345, Grichener_2019}. {As shown by \citet{Hotokezaka_2018}, combining the Galactic \rp\ mass budget, abundance constraints from \textit{r}-process-enhanced metal-poor stars, dwarf galaxy enrichment, and geological measurements favors \rp\ sources with a relatively high per-event yield and a correspondingly low event rate. However, this constraint does not by itself exclude any of the proposed sites.}
    Among these candidates, BNS mergers are actively studied channels and are naturally expected to occur with a delay relative to star formation, which is due to the stellar evolution (stellar birth to compact object binary formation) and GW inspiral time until merger \citep{1964PhDT........51P}. Historically, this delay has been modeled with a power-law DTD using a power-law index, $\alpha$, and a minimum delay time cutoff, $t_\mr{min}$ \citep{1992ApJ...389L..45P, Safarzadeh_2019, McCarthy_2020}.

    In this work, we focus on BNS mergers as the only \textit{r}-process enrichment channel.
    Following \citet{saleem2026}, we divide the BNS into two subpopulations. The first is a delayed channel, labeled ``\textit{Delayed}'', which represents mergers occurring after a time delay relative to star formation. The second is a fast-merging channel, labeled ``\textit{FM}'', which is assumed to trace star formation with no delay. The inclusion of such a prompt component is motivated by some  Galactic neutron star and sGRB studies that suggest the existence of a rapidly merging subpopulation \citep{Beniamini_2019, Galaudage_2021,beniamini2024, pracchia2026shortgammarayburstprogenitors}.

\subsection{Galactic Chemical Evolution}
\label{sec:gce}
    The Galactic chemical evolution framework is built from the one-zone model described in \cite{Chen_2025} originating from \cite{Siegel_2019_1} (see more details in Appendix \ref{sec: GCE}). 
    
    At each time in cosmic history, our model assumes the Milky Way is enriched by the astrophysical events according to the product of the event rate density at that cosmic time, $R(z)$, with the average event yield, $\overline{m}_\mr{ej}(z)$, 
    \begin{equation}\label{eq:product}
    \mathcal{Y}(z) = R(z)\times\overline{m}_\mr{ej}(z).
    \end{equation}
    This is the total mass of each element injected into the ISM per unit time per unit volume (in $M_\odot\,\mr{Gpc^{-3}\,yr^{-1}}$), or the enrichment efficiency. As a result, any combination of $R(z)$ and $\overline{m}_\mr{ej}(z)$ that preserves this product for each chemical element produces an identical \textrm{[Eu/Fe]}--\textrm{[Fe/H]} track (Figures~\ref{fig:all_track_Eu} and ~\ref{fig:all_track_enrichment}), rendering the two parameters fundamentally degenerate.

    For europium, the Eu yield per event is derived by rescaling the total \textit{r}-process ejecta mass by the solar \textit{r}-process mass fraction for atomic mass number $A \geq 69$ \citep{Arnould_2007}. This is because, for the high metallicity stars we consider, the ISM has been enriched by a large number of events. The nearly universal abundance pattern observed in metal-poor stars \citep{frebel2023observationsrprocessstarsmilky} that follows the solar \textit{r}-process residual serves as a good approximation for the average over multiple events.

Throughout this study, we use the cosmic SFH from \citet{Madau_2017}. However, we also explore the impact of different SFHs, and do not find any qualitative change in our conclusions (see more details in Discussion). We adopt cosmological parameters $\Omega_{\mr{M}} = 0.3$, $\Omega_{\Lambda} = 0.7$, $\Omega_{\mr{b}} = 0.046$, and $h=0.7$.

\subsection{Evolving Enrichment Efficiency}
    The main extension explored in this work is the possibility that the BNS \rp\ enrichment efficiency evolves with redshift in addition to SFH.       
    To model this evolution, we adopt a broken power-law parameterization in $(1+z)$,
    \begin{equation}
    \epsilon(z)
    =
    \begin{cases}
    (1+z)^{\beta}, & z < z^\star, \\[6pt]
    \dfrac{(1+z^\star)^\beta}{(1+z^\star)^\delta}(1+z)^{\delta}, & z \ge z^\star,
    \end{cases}
    \label{eq:efficiency}
    \end{equation}
    $\beta$ controls the low-redshift evolution, $\delta$ controls the high-redshift evolution, and $z^\star$ is the transition redshift. The prefactor in the second branch ensures continuity. When there is no evolution, $\epsilon(z)=1$. We adopt a broken power law because it flexibly accommodates increasing or decreasing formation efficiencies and a transition between the two behaviors. Although our inferred $z^\star$ is weakly constrained, it lies outside the valid range of the one-zone model and therefore does not affect our conclusions.

    For the BNS population assumed to follow the SFH with negligible delay (``\textit{FM}''), the \rp\ enrichment efficiency is modeled as
    \begin{equation}
    \mathcal{Y}_{FM} (z)=\psi(z)\,\eta_0\,\epsilon(z)\,\overline{m}_\mr{ej,0}, 
    \label{eq:fm rate}
    \end{equation}
  where $\psi(z)$ is the cosmic star formation rate density, $\eta_0$ is the local ($z=0$) formation efficiency of the BNS mergers, and $\overline{m}_\mr{ej, 0}$ represents the local yield.
The BNS formation efficiency is defined as the number of BNS mergers that merge within Hubble time per stellar mass formed, ${\mr{d}N_\mr{BNS}}/{\mr{d}M_{\mr{SFR}}}.$

For the BNS population delayed with respect to star formation (``\textit{Delayed}''), the \rp\ enrichment efficiency is given as
    \begin{equation}
    \begin{aligned}
    \mathcal{Y}_{Delayed}(z)
    &= \\
    &\hspace{-5.0em}\int_z^\infty 
    p\!\left( t_{\rm delay} \mid \alpha, t_{\rm min} \right)
    \psi(z')\,\eta_0\,\epsilon(z')\,\overline{m}_\mr{ej, 0} \,
    \dv{t}{z'}\,\mr{d}z',
    \end{aligned}
    \label{eq:delayed rate}
    \end{equation}
    where $t$ denotes the cosmological look-back time; $\alpha$ and $t_{\rm min}$ are the DTD parameters. $z'$ is the redshift of star formation and $z$ is the redshift of the merger, therefore, the delay time $t_\mr{delay}$ is defined as
    $
        t_\mr{delay} = t(z') - t(z).
    $

\label{subsec: model}

\subsection{Bayesian Inference Framework}
To properly propagate observational or theoretical uncertainties, we utilize a Bayesian framework employed in \citet{Chen_2025} and \citet{saleem2026}, with additional evolution of enrichment.
    
    We analyze the stellar abundance measurements of disk stars reported by \citet{Battistini_2016} and infer the posterior distributions of the parameters that govern our chemical evolution models.   
    Each stellar observation is represented as a two-dimensional Gaussian centered on the measured abundance values, with the observational uncertainties defining the covariance matrix. 
        For a given set of parameters, the chemical evolution model predicts a trajectory in the $([\mr{Fe/H}], [\mr{Eu/Fe}])$ plane.
    We construct the likelihood by integrating this Gaussian along the model track for each star, and then taking the product over all observed stars. Thus, the likelihood is given as
    \begin{equation}
        \mathcal{L}(d \mid \vec{\theta},\mathcal{M}) \propto \prod_{i=1}^{N_\mr{disk}}\int \mathcal{N}(\mr{[Fe/H]}, \mr{[Eu/Fe]} \mid d_i)\mr{d}[\mr{Fe/H}],
        \label{likelihood}
    \end{equation}
    where $\mathcal{N}$ denotes the 2D Gaussian for each star. We use Bayesian analysis to infer parameters that control the \textit{r}-process channel, including the local BNS merger rate, local average ejecta mass ($\overline{m}_\mr{ej,0}$), DTD parameters, and the enrichment evolution parameters $\{\beta, \delta, z^\star\}$ in Eq.~\eqref{eq:efficiency}.  We also use the local rate ratio of the \textit{FM} population, $f_\mr{FM}$, to track the fraction of the FM channel relative to the local BNS merger rate.

    For BNS mergers, we utilize a joint prior for the local BNS merger rate $R_0$ and \rp\ ejecta mass informed by Galactic pulsar and LIGO--Virgo--KAGRA (LVK) GWTC-4.0 observations using a semi-analytical fit from numerical relativity simulations. 
    We account for uncertainty in the ejecta mass estimate following \citep{Chen_2025}. We calculate $m_{\rm dyn}$ and $m_{\rm disk}$ from \citep{Kr_ger_2020}, and estimate the total local ejecta mass as

    $$
    m_{\rm ej,0}=\alpha_{\rm dyn}m_{\rm dyn}+f_{\rm loss}m_{\rm disk}.
    $$
    Here, $\alpha_{\rm dyn}\sim U(0.5,1.5)$ represents an adopted \(50\%\) uncertainty in the dynamical-ejecta estimate, while $f_{\rm loss}\sim U(0.15,1)$ accounts for uncertainty in the fraction of the mass ejected from the post-merger disk. Both quantities are sampled when constructing the joint prior $p(R_0,\overline{m}_{\rm ej,0})$, so the uncertainties are propagated through Bayesian inference.

    We construct our prior distribution for DTD parameters using the posterior samples of sGRB observations reported in \cite{Zevin_2022}. The priors for the enrichment efficiency parameters are uniform. We summarize the adopted prior distributions in Table~\ref{tab:priors} in the Appendix.
    Parameter inference is performed with \texttt{bilby} \citep{bilby_paper} using the \texttt{Dynesty} \citep{Speagle_2020} nested sampler.

\section{Results}
\label{sec:results}
    We explore nine scenarios for the production of \textit{r}-process elements (Table~\ref{tab:models}). The models include different combinations of \textit{Delayed} and \textit{FM} channels, with or without evolution. $\mathcal{M}_1$--$\mathcal{M}_4$ correspond to single-channel scenarios, while $\mathcal{M}_5$--$\mathcal{M}_9$ correspond to double-channel scenarios. 
    
    The total local BNS merger rate is kept consistent with the GWTC--4.0 Full-Pop local BNS merger rate $89^{+159}_{-67}\mr{Gpc^{-3}}\;\mr{yr^{-1}}$ \citep{theligoscientificcollaboration2025gwtc40populationpropertiesmerging}, treating \textit{Delayed} and \textit{FM} channels as subpopulations of BNSs, assuming both channels detectable in current GW observations, and sharing the same mass distribution. Note that scenarios $\mathcal{M}_1$ and $\mathcal{M}_5$ are the same scenarios studied in \cite{Chen_2025} and \cite{saleem2026}, respectively.

    {The Bayesian inference results are shown in Table~\ref{table: results}, which includes the median of the parameters and the 90\% interval. }

    \begin{table}
        \centering
        \caption{Summary of the nine scenarios considered in this work, categorized by whether the \textit{Delayed} and \textit{FM} BNS channels are included and whether their enrichment efficiencies are held constant or allowed to evolve beyond SFH. $\mathcal{M}_1-\mathcal{M}_4$ are single-channel scenarios, either \textit{Delayed} or \textit{FM} as the only \textit{r}-process producing channel. $\mathcal{M}_5-\mathcal{M}_9$ are double-channel scenarios.}
        \begin{tabular}{lcc}
             
            \toprule
            Scenario & Delayed BNS & Fast-Merging BNS \\
            \midrule
            $\mathcal{M}_1$    & non-evolving        & $\times$ \\
            $\mathcal{M}_2$    & evolving            & $\times$ \\
            $\mathcal{M}_3$    & $\times$            & non-evolving \\
            $\mathcal{M}_4$    & $\times$            & evolving \\
            $\mathcal{M}_5$    & non-evolving        & non-evolving \\
            $\mathcal{M}_6$    & non-evolving        & evolving \\
            $\mathcal{M}_7$    & evolving            & non-evolving \\
            $\mathcal{M}_8$    & \multicolumn{2}{c}{evolving independently} \\
            $\mathcal{M}_{9}$ & \multicolumn{2}{c}{evolving identically} \\
            \bottomrule
        \end{tabular}
        
        \label{tab:models}
    \end{table}

    \begin{deluxetable*}{lccccccccccc}
    \label{table: results}
    \tablecaption{Median and the 90\% credible intervals of the parameter inference. 
    \label{tab:parameter_value}}
\tablehead{
      \colhead{Scenario} & 
      \colhead{Local Rate $R_0$ [$\mr{Gpc^{-3}\,yr^{-1}}$]} & 
      \colhead{$\overline{m}_\mr{ej}$ [$M_\odot$]} & 
      \colhead{$\alpha$} &
      \colhead{$t_\mr{min}\,\mr{[Gyr]}$} &
      \colhead{$\beta_{Delayed}$} & 
      \colhead{$\delta_{Delayed}$} & 
      \colhead{$z^\star_{Delayed}$} & 
      \colhead{$\beta_{FM}$} & 
      \colhead{$\delta_{FM}$} & 
      \colhead{$z^\star_{FM}$} & 
      \colhead{$f_{FM}$}
    }
    \startdata
    $\mathcal{M}_1$ & $123.1^{+123.4}_{-45.5}$ & $0.07^{+0.04}_{-0.03}$ & $-1.98^{+0.11}_{-0.12}$ & $0.024^{+0.001}_{-0.002}$ & N/A & N/A & N/A & N/A & N/A & N/A & N/A \\
    $\mathcal{M}_2$ & $87.0^{+80.9}_{-37.8}$ & $0.04^{+0.03}_{-0.02}$ & $-1.77^{+0.31}_{-0.33}$ & $0.18^{+0.07}_{-0.07}$ & $1.8^{+0.5}_{-0.4}$ & $-3.3^{+7.7}_{-6.0}$ & $5.9^{+3.6}_{-2.2}$ & N/A & N/A & N/A & N/A \\
    $\mathcal{M}_3$ & $122.4^{+126.6}_{-47.5}$ & $0.06^{+0.04}_{-0.03}$ & N/A & N/A & N/A & N/A & N/A & N/A & N/A & N/A & N/A \\
    $\mathcal{M}_4$ & $90.5^{+85.5}_{-37.5}$ & $0.04^{+0.03}_{-0.02}$ & N/A & N/A & N/A & N/A & N/A & $0.8^{+0.2}_{-0.2}$ & $-4.6^{+4.7}_{-4.9}$ & $3.9^{+3.2}_{-1.3}$ & N/A \\
    $\mathcal{M}_5$ & $119.3^{+112.4}_{-41.9}$ & $0.07^{+0.04}_{-0.03}$ & $-1.80^{+0.35}_{-0.38}$ & $0.18^{+0.07}_{-0.08}$ & N/A & N/A & N/A & N/A & N/A & N/A & $0.99^{+0.01}_{-0.03}$ \\
    $\mathcal{M}_6$ & $91.9^{+83.2}_{-39.8}$ & $0.04^{+0.04}_{-0.02}$ & $-1.82^{+0.35}_{-0.36}$ & $0.19^{+0.07}_{-0.08}$ & N/A & N/A & N/A & $1.0^{+0.6}_{-0.3}$ & $-3.6^{+3.6}_{-5.6}$ & $3.7^{+2.0}_{-1.3}$ & $0.7^{+0.3}_{-0.4}$ \\
    $\mathcal{M}_7$ & $91.8^{+92.5}_{-40.5}$ & $0.04^{+0.04}_{-0.02}$ & $-1.83^{+0.32}_{-0.35}$ & $0.19^{+0.06}_{-0.08}$ & $2.2^{+1.6}_{-0.6}$ & $-3.3^{+5.7}_{-6.0}$ & $4.8^{+4.2}_{-2.1}$ & N/A & N/A & N/A & $0.5^{+0.3}_{-0.5}$ \\
    $\mathcal{M}_8$ & $103.4^{+98.2}_{-47.3}$ & $0.05^{+0.04}_{-0.03}$ & $-1.85^{+0.29}_{-0.36}$ & $0.19^{+0.06}_{-0.07}$ & $1.8^{+0.6}_{-0.4}$ & $-3.4^{+7.4}_{-5.8}$ & $5.5^{+3.7}_{-2.1}$ & $-4.8^{+5.5}_{-4.5}$ & $-0.9^{+9.4}_{-8.1}$ & $5.3^{+4.2}_{-4.2}$ & $0.4^{+0.3}_{-0.4}$ \\
    $\mathcal{M}_9$ & $89.1^{+72.4}_{-38.9}$ & $0.04^{+0.03}_{-0.02}$ & $-1.83^{+0.35}_{-0.38}$ & $0.18^{+0.07}_{-0.08}$ & $1.1^{+0.5}_{-0.4}$ & $-4.2^{+6.3}_{-5.3}$ & $4.4^{+4.4}_{-1.4}$ & N/A & N/A & N/A & $0.5^{+0.5}_{-0.4}$ \\
    \enddata
    \tablecomments{N/A indicates that the parameter is not included in the given scenario.}
    \end{deluxetable*}

\subsection{Evolving vs. Non-evolving}
\label{sub: Evolving vs. Non-evolving}
\begin{figure}
        \centering
        \includegraphics[width=0.5\textwidth]{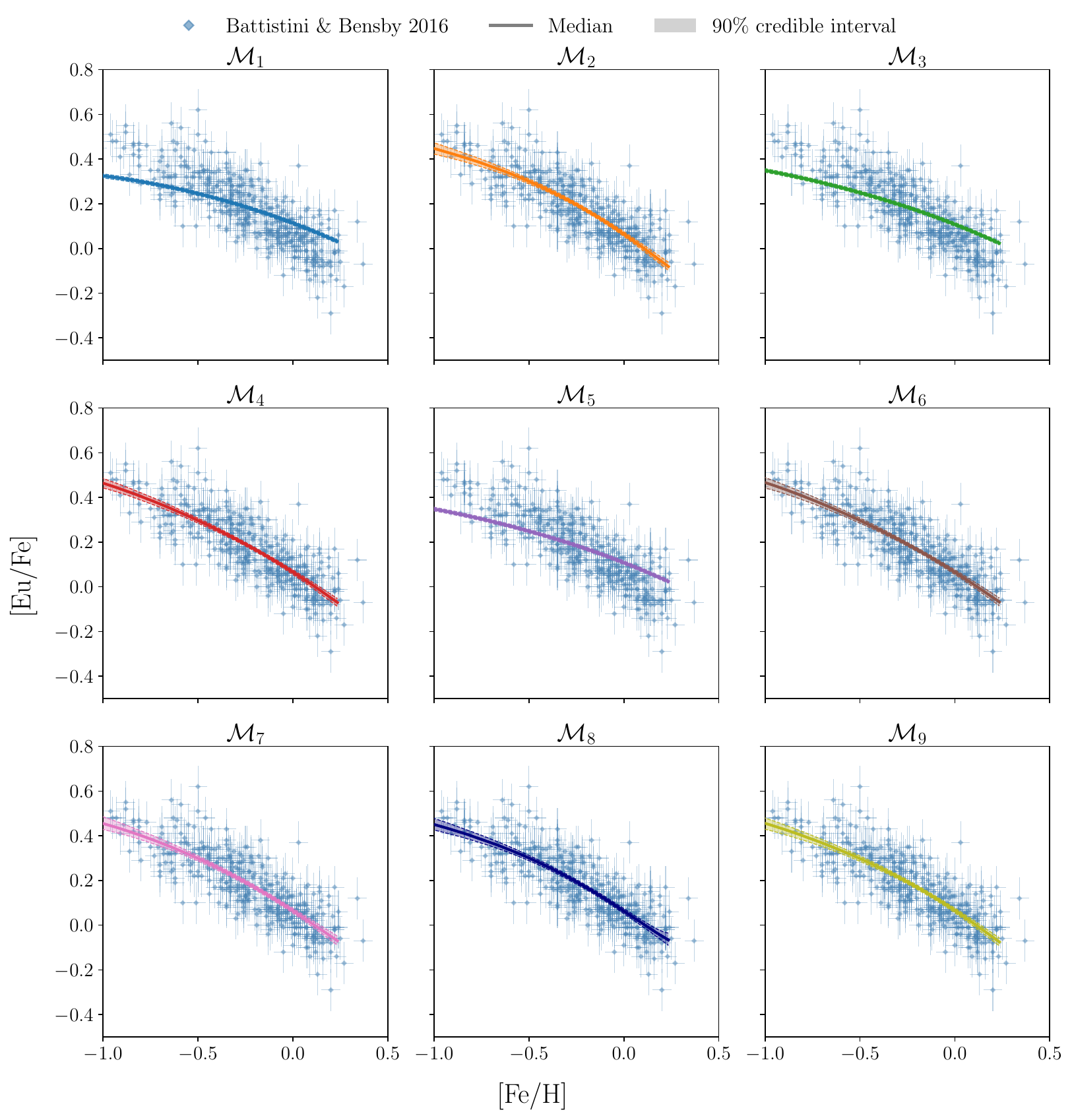}
        \caption{Median and $90\%$ credible interval of the inferred $\mathrm{[Eu/Fe]}$--$\mathrm{[Fe/H]}$ tracks of all of the scenarios compared against disk-star abundance measurements from \cite{Battistini_2016}.}
        \label{fig:all_track_Eu}
    \end{figure}

\begin{figure}
        \centering
        \includegraphics[width=0.5\textwidth]{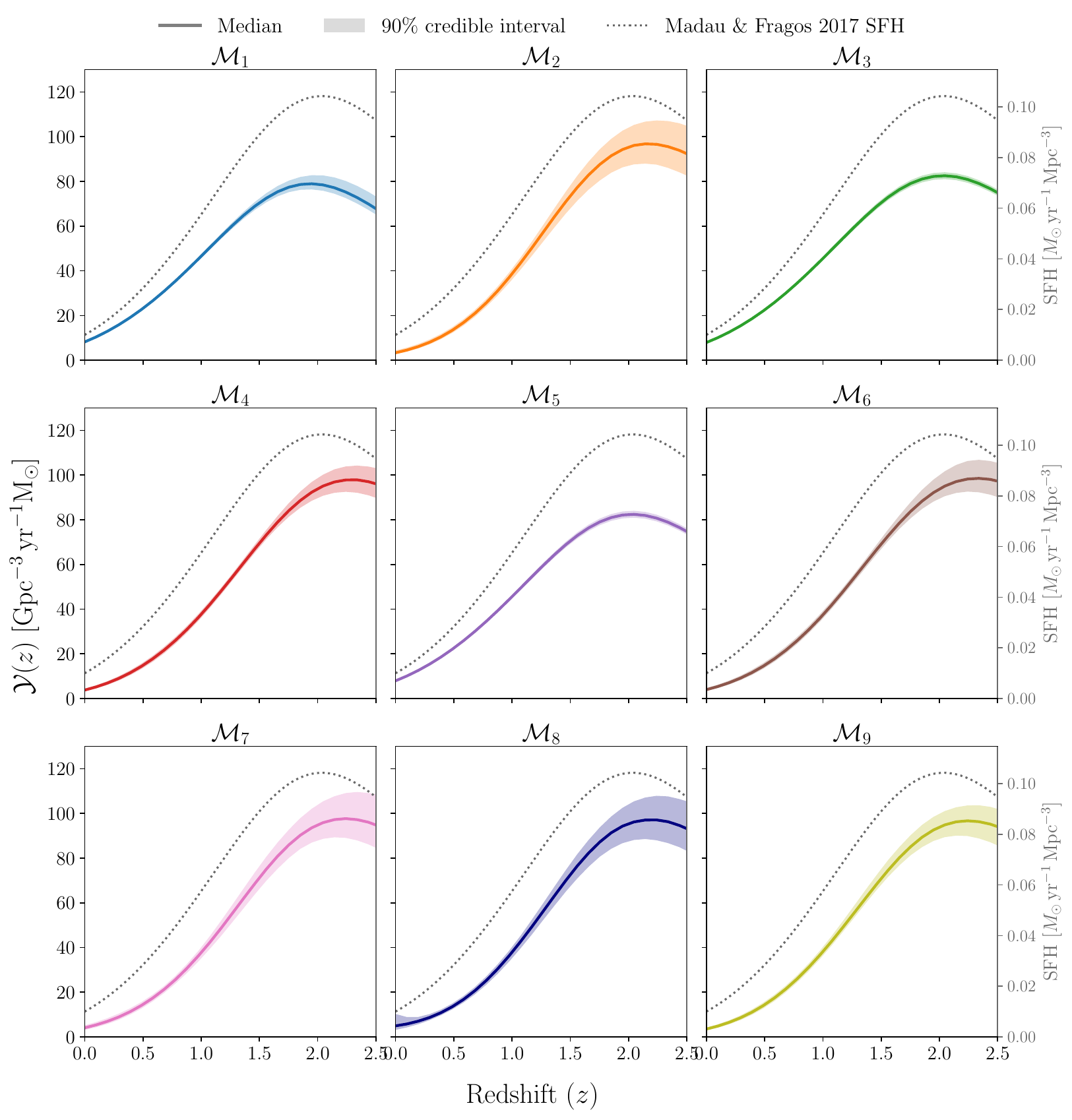}
        \caption{Median and $90\%$ credible interval of the inferred overall \textit{r}-process enrichment efficiency of all of the scenarios as a function of redshift (Eq.~\ref{eq:product}). We also overplot the SFH used in this analysis from \citet{Madau_2017} for comparison of the peak location (black dotted line).}
        \label{fig:all_track_enrichment}
    \end{figure}

    We first compare the stellar abundance tracks inferred by evolving and non-evolving scenarios. 
    Figure~\ref{fig:all_track_Eu} is the median and $90\%$ credible interval of inferred stellar abundance track. We find that all scenarios with an evolving enrichment efficiency describe the Milky Way disk-star abundance data \citep{Battistini_2016} remarkably better than the non-evolving scenarios. This improvement is driven by the timing of \textit{r}-process enrichment: reproducing the negative slope in the of Figure~\ref{fig:all_track_Eu} requires an enhanced early contribution from BNS mergers, prior to the peak of iron production by CCSNe (which strictly follows the SFH), as shown in Figure~\ref{fig:all_track_enrichment}.

    $\mathcal{M}_{1}$ represents the scenario where the \textit{Delayed} channel is the only source of \textit{r}-process production, which cannot reproduce the disk star abundance patterns, as explored in previous work (e.g., \citealt{Chen_2025} and \citealt{molero2025}).
    On the other hand, $\mathcal{M}_2$ allows the \textit{Delayed} channel to evolve, and fully reproduces the stellar abundance data. The Bayes factor between $\mathcal{M}_2$ and $\mathcal{M}_1$, $B^2_1$, is larger than $10^{40}$ (see Appendix \ref{subsec: Evidence} and Table~\ref{tab:log-evidence}). 

    {$\mathcal{M}_3$ and $\mathcal{M}_4$ treat the \textit{FM} channel as the sole $r$-process source, without and with evolution, respectively. While $\mathcal{M}_3$ fails to fully reproduce the observed abundance trend even with SFH-tracking early enrichment, $\mathcal{M}_4$ succeeds once the enrichment efficiency is allowed to increase toward high redshift ($\beta = 1.8^{+0.5}_{-0.4}$). This result again suggests that Eu enrichment must occur on timescales earlier than CCSNe, whereas following Eu injection must be suppressed relative to the early contribution. In this sense, the inferred evolution in $\mathcal{M}_4$ may be interpreted as a smooth suppression of prompt \rp\ production toward low redshift. We note that such behavior is also qualitatively similar to the metallicity-dependent formation decrease expected for collapsars, for which GRB-producing events are observed predominantly in low-metallicity environments and become increasingly rare above a characteristic metallicity threshold below solar \citep{2005A&A...443..643Y,2006ApJ...637..914W}. A related scenario was explored by \citet{Tarumi_2021}, who modeled the collapsar contribution with an additional evolution relative to the SFH, effectively corresponding to a ``negative'' delay time, and tested whether such enhanced early enrichment could reproduce the observed \rp\ abundances of metal-poor stars.}

    $\mathcal{M}_{5}$ is the scenario investigated in \cite{saleem2026}, where both \textit{Delayed} and \textit{FM} channels were taken into account, however, without any evolution. They conclude that in order to explain the disk star abundance pattern, the \textit{FM} rate density has to be more than two orders of magnitude higher than the \textit{Delayed} rate density across all redshifts.
    Even in this extreme scenario, the inferred $\mr{[Eu/Fe]}-\mr{[Fe/H]}$ trend fails to fully recover the observations (Figure \ref{fig:all_track_enrichment}). In addition, existing theories do not predict the \textit{FM} channel to dominate at close to $100\%$ of the total population \citep{Dominik_2012, belczynski2018binaryneutronstarformation}.
    Motivated by this result, we explore a range of combinations of double channel scenarios with evolution, and find that all scenarios with evolution show decisive Bayes evidence over $\mathcal{M}_{5}$, with Bayes factors exceeding $10^{23}$ (See Table~\ref{tab:log-evidence} in Appendix). 

    We note that $\mathcal{M}_8$ allows the \textit{Delayed} and \textit{FM} channels to evolve independently. However, because of its large number of parameters, the inference is not well constrained. In contrast, $\mathcal{M}_9$ requires the \textit{Delayed} and \textit{FM} channels to share the same evolution parameters, $\beta_{\mr{Delayed}}$, $\delta_{\mr{Delayed}}$, and $z^{\star}_{\mr{Delayed}}$, resulting in a more constrained inference.

    Finally, when comparing single- versus double-channel with evolution, we find comparable Bayes factors among these scenarios. Either the \textit{Delayed} or \textit{FM} channel alone, as well as a combination of both, can reproduce the stellar abundance data equally well when evolution is included.

\subsection{Evolution in Rate}
    The evolution of both the BNS merger rate and yield could contribute to the evolution of the overall \textit{r}-process enrichment efficiency. 
    To evaluate whether such evolution is possible, we first assume that the entire evolution arises from the evolution of the merger rate and compare the rate with observations and theoretical predictions in this subsection. In the next subsection, we explore the possibility of yield evolution. These comparisons are presented separately due to the rate and yield degeneracy.
\subsubsection{Comparison with sGRB Rates}

\begin{figure}
    \centering
    \includegraphics[width=1.0\linewidth]{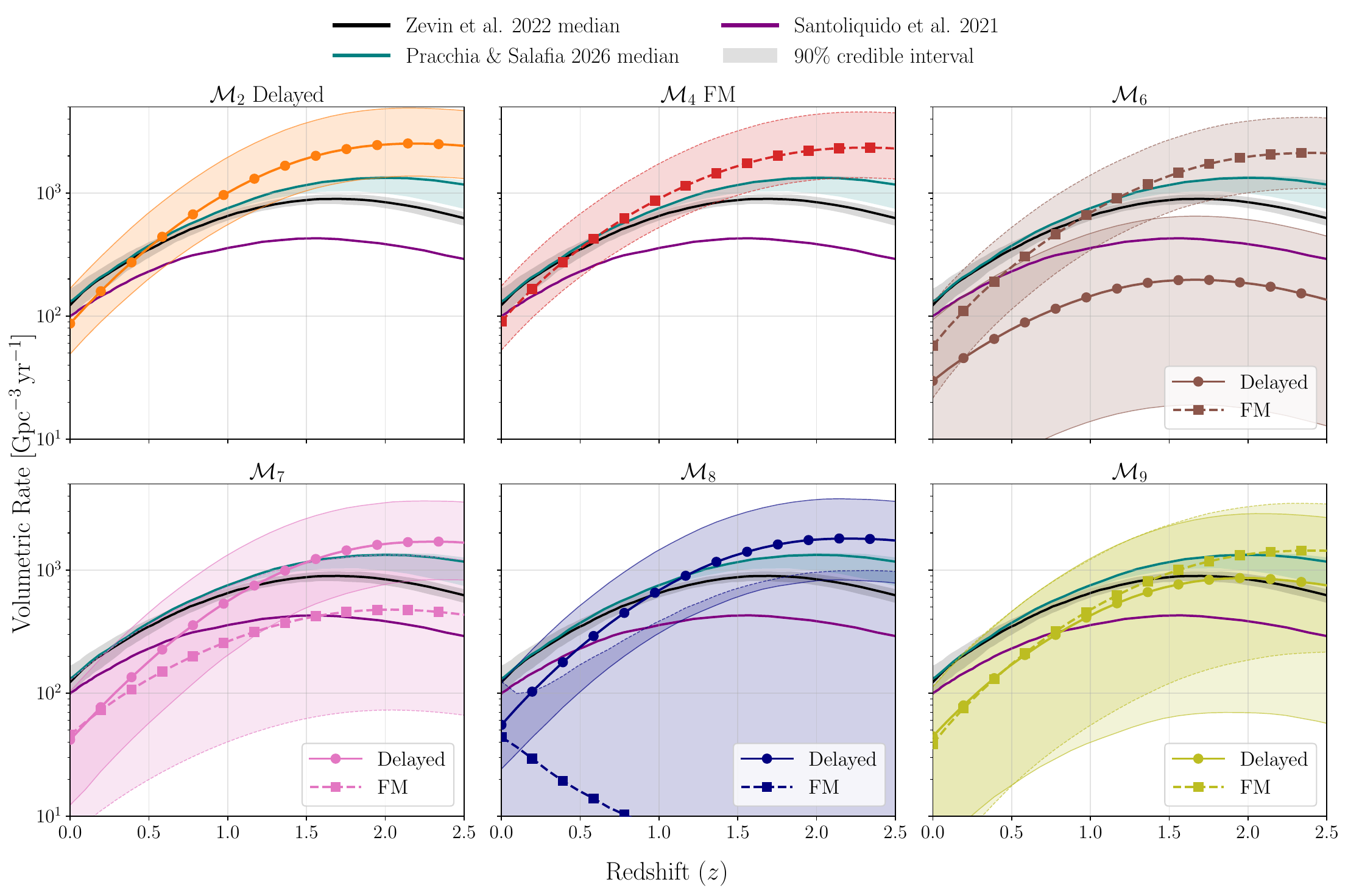}
    \caption{Comoving BNS merger rate as a function of redshift for scenarios with evolution. Circles represent the \textit{Delayed} population, while squares represent the \textit{FM} population. The black and teal curves show the median sGRB rates inferred by \citet{Zevin_2022} and \citet{pracchia2026shortgammarayburstprogenitors}, respectively, with the corresponding shaded regions showing their $90\%$ credible intervals. For \citet{pracchia2026shortgammarayburstprogenitors}, we show the power-law empirical luminosity-function model. The purple line shows the BNS merger rate predicted by the population-synthesis study of \citet{Santoliquido_2021}, normalized to a local rate of $100\,\mathrm{Gpc^{-3}\,yr^{-1}}$. For the single-channel evolving scenarios ($\mathcal{M}_2$, $\mathcal{M}_4$), the local rates agree with the sGRB constraints but the median rates rise more rapidly at $z\gtrsim 1$. For the double-channel evolving scenarios ($\mathcal{M}_6$, $\mathcal{M}_7$, $\mathcal{M}_8$, $\mathcal{M}_9$), the local rates lie on the edge of the $90\%$ credible interval, and the tension at high redshift persists.}
    \label{fig:sgrb}
    \end{figure}

    BNS mergers are known progenitors of sGRBs \citep{Abbott_2017, 
    Goldstein_2017, Nicholl_2017} (see, however, possible tension between their rates; e.g., \citealt{2018MNRAS.477.4275P, 2026Fishbach}). In Figure~\ref{fig:sgrb}, we compare the merger rate predicted by our evolving scenarios with those inferred from sGRB observations, showing the median and $90\%$ credible intervals of the rate as a function of redshift.
    
    A range of DTD parameters has been inferred in the sGRB literature. Earlier studies found moderately steep power-law indices of approximately $\alpha\sim-1$ using sGRB observations \citep{Wanderman_2015} and Galactic binary-pulsar observations \citep{Beniamini_2019}. Among more recent studies, \citet{Zevin_2022} used the reconstructed SFHs of sGRB host galaxies to infer a steeper DTD, with $\alpha=-1.83^{+0.35}_{-0.39}$ and a minimum delay time of $t_{\min}=184^{+67}_{-79}\,\mathrm{Myr}$. In contrast, \citet{pracchia2026shortgammarayburstprogenitors} analyzed the sGRB observations within a hierarchical framework and found the population to be consistent with predominantly short delay times, with average delays spanning approximately $10$--$800\,\mathrm{Myr}$. Shorter delays favored by \citet{pracchia2026shortgammarayburstprogenitors} produce a merger history that more closely follows the cosmic SFH, whereas the rate inferred by \citet{Zevin_2022} is shifted toward later cosmic times. These constraints are also subject to different systematic uncertainties associated with host-galaxy SFHs, sample selection, luminosity functions, and the assumed sGRB population properties.
    
    As shown in Figure~\ref{fig:sgrb}, the behavior of the evolving scenarios depends on the channel composition. For the single-channel evolving scenarios ($\mathcal{M}_2$ and $\mathcal{M}_4$), the local rates are broadly consistent with the sGRB constraints, but the rates rise more rapidly toward high redshift ($z\gtrsim 1$) than either sGRB-inferred rate, even when the $90\%$ credible intervals are taken into account. For the double-channel evolving scenarios ($\mathcal{M}_6$, $\mathcal{M}_7$, $\mathcal{M}_8$, $\mathcal{M}_9$), the local rates lie on the edge of the $90\%$ credible interval of the sGRB constraints, and the tension at high 
    redshift persists.
    
    The comparison with \citet{Zevin_2022} is not fully independent for scenarios containing the \textit{Delayed} channel, because their DTD posterior is adopted as our prior on $(\alpha, t_{\min})$. Since our recovered DTD parameters remain close to this prior, the discrepancy in the predicted rate histories is driven primarily by the redshift-dependent formation efficiency required by the Milky Way abundance data, rather than a departure in the DTD itself. In contrast, the \textit{FM} channel carries no prior informed by either sGRB study and the broken power-law parameterization is flexible enough to reproduce various rate evolutions, including one that closely follows the cosmic SFH as favored by \citet{pracchia2026shortgammarayburstprogenitors}. The comparison between the \textit{FM} evolving scenarios and the sGRB constraints is therefore more independent, and the tension in these scenarios provides additional support for our conclusion.
    
\subsubsection{Comparison with the Stochastic GW Background}\label{sec:sgwb-comparison}
    An independent consistency check on the merger-rate histories inferred in this work is whether they remain compatible with existing gravitational-wave constraints. In particular, enhanced merger activity at high redshift can increase the stochastic gravitational-wave background (SGWB) expected from BNS mergers. We therefore compute the SGWB implied by each merger-rate evolution scenario and compare it with the latest LVK upper limits from the O1--O4a observing runs~\citep{theligoscientificcollaboration2025upperlimitsisotropicgravitationalwave}.

    For an inspiral-dominated population of quasi-circular compact binaries, the SGWB in the detector band follows the standard power-law form $\Omega_{\rm GW}(f)=\Omega_0(f/f_0)^{2/3}$, with the normalization $\Omega_0$ determined by the underlying merger-rate evolution ~\citep{Phinney:2001di,Christensen_2018}. We use $f_0=25\,\mr{Hz}$ as the reference frequency. Details of the SGWB calculation are given in Appendix~\ref{app:sgwb}.

\begin{figure}
        \centering
        \includegraphics[width=1.0\linewidth]{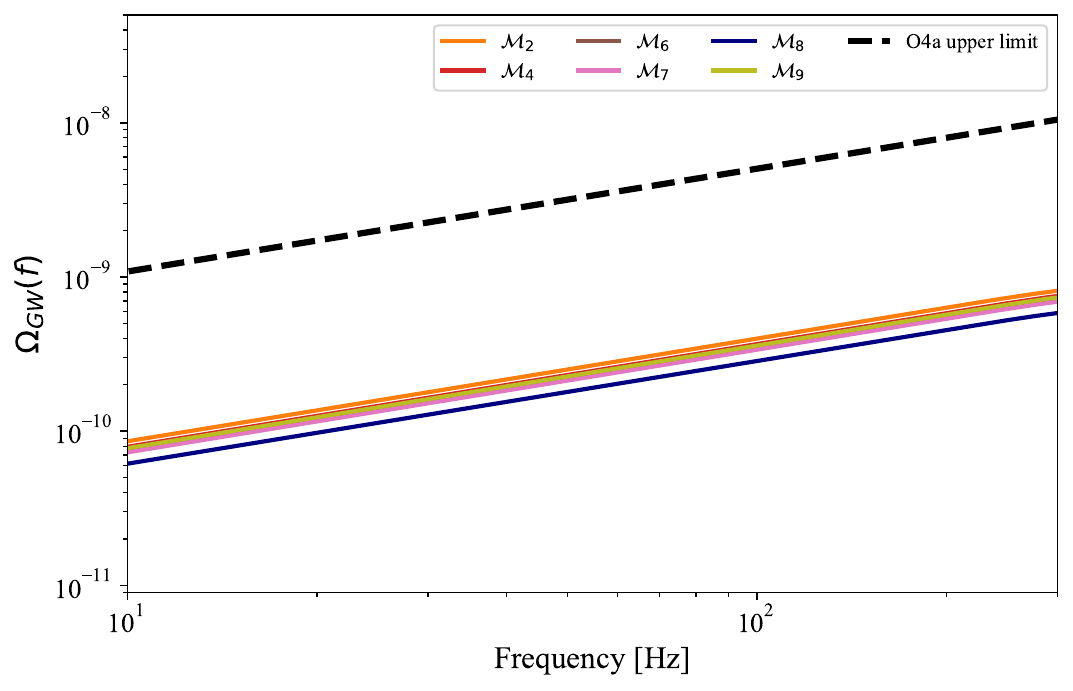}
        \caption{Predicted stochastic gravitational-wave background energy-density spectra, $\Omega_{\rm GW}(f)$, for scenarios with evolution. Each curve is computed using the corresponding inferred BNS merger rate history under an inspiral-only equal-mass $1.4+1.4\,M_\odot$ approximation. All spectra exhibit the characteristic $\Omega_{\rm GW}(f)\propto f^{2/3}$ scaling across the plotted band. For comparison, we show the LVK O4a upper limit on the SGWB with a power-law index of 2/3 (black dashed line; \citealt{theligoscientificcollaboration2025upperlimitsisotropicgravitationalwave}). 
        All scenarios remain below the current observational constraint over the expected inspiral frequency range.}
        \label{fig:powerspectrum}
    \end{figure}

    We find that all merger-rate evolution scenarios considered in this work, including those with the strongest high-redshift enhancement, remain below the current LVK upper limits across the frequency band considered here (Figure~\ref{fig:powerspectrum}). Thus, the early-merging BNS populations inferred in this work do not overproduce the SGWB relative to present detector sensitivities.

\subsubsection{Comparison with Population Synthesis Studies}
We compare the inferred BNS merger rate with predictions from population synthesis studies. The merger rate inferred under scenarios with evolution appears to rise more rapidly at higher redshifts than multiple population synthesis predictions \citep{van_Son_2025, Broekgaarden_2022, Santoliquido_2021, Neijssel_2019}. Figure~\ref{fig:sgrb} shows a comparison with the rate predicted by \cite{Santoliquido_2021}. 

\begin{figure}
        \centering
        \includegraphics[width=1.0\linewidth]{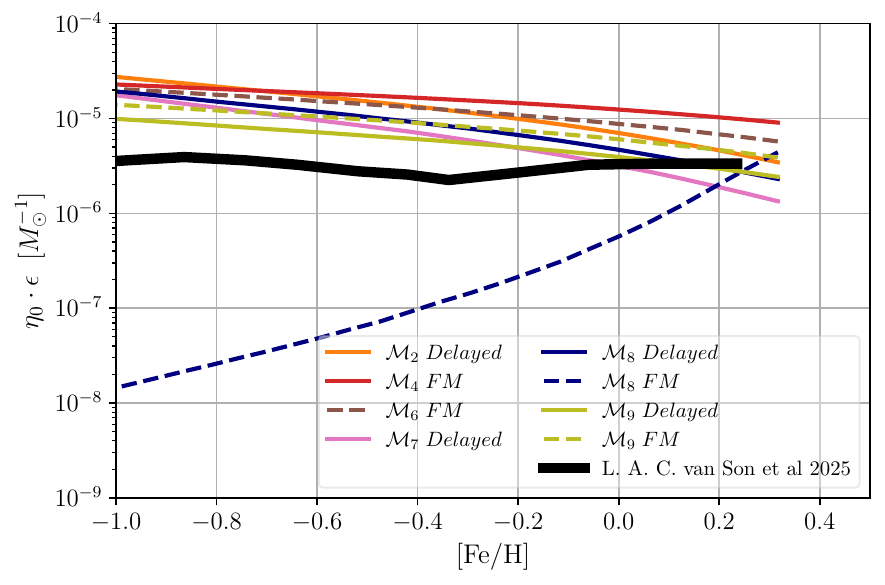} 
        \caption{Median of the BNS formation efficiency $\eta_0\cdot \epsilon$ as a function of [Fe/H]. We also plot the median trend summarized in \cite{van_Son_2025}, which shows no correlation between formation efficiency and metallicity, in contrast to the scenarios explored here.
        }
        \label{fig: all_eff}
    \end{figure}

    We also compare the BNS formation efficiency as a function of the metallicity. \cite{van_Son_2025} summarized results from several population synthesis studies, which generally find no strong correlation between BNS formation efficiency and metallicity. We calculate the formation efficiency as
 $$\cfrac{\mr{d}N_\mr{BNS}}{\mr{d}M_{\mr{SFR}}}=\eta_0 \cdot \epsilon(z).$$
    and compare it to the trend summarized by \cite{van_Son_2025} in Figure~\ref{fig: all_eff}. Scenarios with evolution indicate a factor of a few to one order-of-magnitude increase in formation efficiency at $\mr{[Fe/H]}\sim -1$, inconsistent with these population synthesis results. On the other hand, there are other studies that suggest possible correlations between BNS formation efficiency and metallicity \citep{gallegosgarcia2023evolutionaryoriginsbinaryneutron,chattaraj2026doubleneutronstardelay}. A direct comparison with the models considered in these studies may help constrain the population synthesis parameters, and therefore the plausible BNS formation history needed to explain observations of stellar abundances in the Milky Way disk.

\subsection{Evolution in yield}
\label{subsec: Evolution of Average Yield}
    In this subsection, we instead explore the possibility that the evolution of the \rp\ enrichment efficiency arises from evolution in the yield.

\begin{figure}
        \centering
        \includegraphics[width=1.0\linewidth]{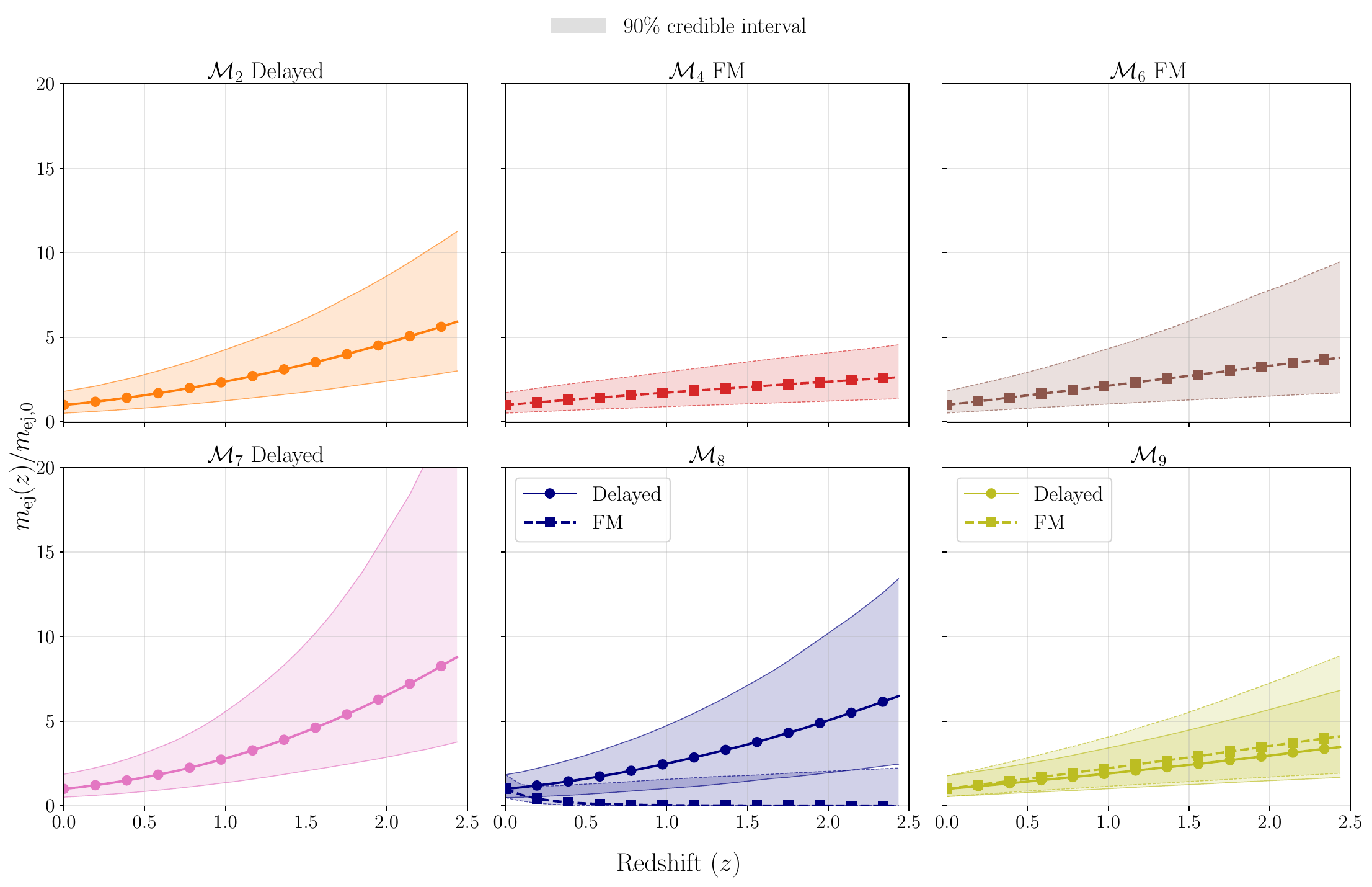}
        \caption{Median and $90\%$ credible interval of average yield as a function of redshift, normalized to the local value. Different scenarios suggest a factor of $2$ to $16$ increase by $z\sim 2$.}
    
        \label{fig:Mej_z}
    \end{figure}

    In Figure~\ref{fig:Mej_z}, we show 
\begin{equation}
\begin{aligned}
\overline{m}_{\mr{ej}}(z)
\equiv
\frac{\mathcal{Y}(z)}{R(z)}
=
\qquad\qquad\qquad\qquad\qquad\qquad\qquad
\\
\begin{cases}
\displaystyle
\mathcal{Y}(z)
\left[
\int_z^\infty
p\!\left(
t_{\rm delay}\mid \alpha,t_{\rm min}
\right)
\psi(z')\,\eta_0\
\dv{t}{z'}
\,\mr{d}z'
\right]^{-1}
&
\mr{Delayed},
\\[2ex]
\displaystyle
\mathcal{Y}(z)[\psi(z)\,\eta_0]^{-1}
&
\mr{FM}
\end{cases}
\end{aligned}
\end{equation}
    normalized by the local value, $\overline{m}_\mr{ej,0}$.
    We find a factor of 2 to 16 increase in the average yield relative to the local value by $z\sim 2$, if all of the evolution is attributed to the yield~\footnote{We note that the BNS merger rate still evolves with redshift following the SFH and DTD under this assumption; it simply does not include any additional redshift evolution.}. Although such evolution could arise from redshift dependence in the BNS mass and spin distributions or from differences in the merger environment in the early Universe, such a dramatic increase in yield poses challenges for BNS formation models and evolutionary history.

\section{Discussion}
\label{sec:discussion}

In this paper, we explore the possibility that the BNS \textit{r}-process enrichment efficiency evolves beyond the SFH. We find that including such additional evolution reproduces the stellar abundance distribution in the Milky Way disk significantly better than scenarios without additional evolution, under the assumption that BNS mergers are the sole progenitors of \textit{r}-process elements.

Both evolution in the BNS merger rate and in the \textit{r}-process yield could contribute, and their effects are degenerate. To evaluate how plausible such evolution is, 
we quantify the required BNS merger rate, the resulting SGWB, and the BNS formation efficiency, and compare them with observations and theoretical predictions. We find that the resulting SGWB remains consistent with LVK observations, while the required merger rate and formation efficiency are in tension with sGRB observations and multiple population synthesis models. 
If the evolution arises entirely from changes in the average yield in the early Universe, we find that at least a factor of two increase in yield by $z\sim 2$ is required. Such a scenario would require highly specific theoretical explanations for the evolution of the BNS merger history and nucleosynthesis yields.

The tensions may be alleviated if both the BNS merger rate and the \rp\ yield evolve simultaneously, although self-consistent theoretical models would still be required, potentially involving a significant degree of fine-tuning, to produce such joint evolution.

The GW detector sensitivities in the next LVK observing run (O5) are expected to improve, with the goal of lowering the stochastic background limits by about  an order of magnitude \citep{theligoscientificcollaboration2025upperlimitsisotropicgravitationalwave}. With this improvement, it may become possible to directly constrain \textit{r}-process enrichment scenarios through comparisons with the SGWB energy-density spectrum (see Figure~\ref{fig:powerspectrum}).

In our study, we find additional evolution in BNS merger rate relative to sGRB observations at higher redshift (Figure~\ref{fig:sgrb}). In parallel, recent studies have also pointed out a potential tension between the local BNS merger rate and the sGRB rate~\citep{2026Fishbach, DeSantis:2026wfw}. These tensions, both in the local Universe and at high redshift (our findings), may indicate that a fraction of sGRBs originate from progenitors other than BNS mergers.

The evolution we find could be interpreted as redshift-dependent BNS DTDs, as explored in~\cite{chattaraj2026doubleneutronstardelay}. A direct comparison may help constrain the population synthesis models required for BNS mergers to serve as the sole progenitors of \textit{r}-process elements.

We note that, while the uncertainty in the BNS yield remains large and may extend beyond the semi-analytical model adopted in this paper, our conclusions regarding the required additional evolution in the BNS merger rate (a different yield model would introduce an offset in Figure~\ref{fig:sgrb} without changing the increasing trend) or in the yield itself (which is already normalized to local value in Figure~\ref{fig:Mej_z}) remain robust. Ultimately, there is only one physically realized, though currently unknown, model for the \textit{r}-process yield.

Although we adopt the cosmic SFH \citep{Madau_2017} in our main analysis, uncertainties in the Milky Way SFH motivate us to examine whether alternative SFH assumptions could affect our conclusions. As an extreme test case, we first assume a constant SFH and find that our main results remain unchanged, including the strong preference for enrichment-evolution scenarios and the comparisons with observational and theoretical constraints on merger rates and yields. We further restrict our data to Milky Way thick-disk stars and repeat the analysis, motivated by the relatively well-constrained SFH of the thick disk~\citep{refId0}. Because the thick disk experienced enhanced star formation at early times and suppressed star formation in more recent epochs, its SFH could, in principle, reduce or even eliminate the need for evolving \rp\ enrichment. However, we again find that this SFH variation cannot reproduce the observed \rp\ abundances without evolution in \rp\ enrichment, leading to the same conclusions as in our fiducial analysis.

In the main analysis, we restrict our treatment to a one-zone chemical evolution model, in which the ISM is assumed to be instantaneously and homogeneously mixed. Consequently, each set of model parameters predicts a single deterministic abundance trajectory, representing the mean evolution of $[\mr{Eu}/\mr{Fe}]$ as a function of $[\mr{Fe}/\mr{H}]$. This approximation does not explicitly capture the star-to-star abundance variations that may arise from inhomogeneous mixing, localized enrichment events, or different mixing histories within the ISM. In Appendix~\ref{sec:scatter}, we therefore extend the likelihood by introducing an intrinsic-scatter parameter around the mean chemical-evolution track. This parameter provides a phenomenological description of the additional abundance dispersion produced by inhomogeneous mixing and allows the inference to accommodate different effective mixing conditions without explicitly adopting a spatially resolved chemical-evolution model. We repeat the full Bayesian inference for each model scenario with this additional parameter, and compute the Bayes factors among the scenarios. Including this additional freedom does not alter the relative preference among the scenarios or any of our conclusions.

{We note that the Bayes factors reported here are conditional on the fixed ingredients of the chemical evolution model (e.g., the SN Ia delay time distribution, the supernova Fe yields, and the star formation history, see \ref{sec: GCE} for further details). Therefore, the uncertainties of such ingredients are not propagated in the evidence calculation. Uncertainties in these ingredients would broaden the predicted [Eu/Fe] tracks and would reduce the magnitude of the Bayes factors, so the values should be interpreted as the relative support for each scenario within this fixed framework.}

In summary, although several alternative \textit{r}-process production candidates have been proposed, all currently face some level of tension with observations or theoretical expectations. In this study, we therefore explore the possibility of additional evolution in the BNS \textit{r}-process enrichment efficiency. By quantifying the required evolution, we are able to directly compare it with observations and theoretical predictions to assess the plausibility of such scenarios. Future observational and theoretical advances will be crucial for determining whether the required evolution can be realized within a self-consistent BNS merger scenario.

\begin{acknowledgments}
The authors would like to thank Alexander Ji and Katerina Chatziioannou for very helpful discussions. M.S.\ is supported by the Weinberg Institute for Theoretical Physics at the University of Texas at Austin. H.-Y.C. is supported by the National Science Foundation under Grant PHY-2308752 and Department of Energy Grant DE-SC0025296. The authors are grateful for computational resources provided by the LIGO Laboratory and supported by National Science Foundation Grants PHY-0757058 and PHY-0823459. This material is based upon work supported by NSF's LIGO Laboratory which is a major facility fully funded by the National Science Foundation.
LLM model Claude was used under full author supervision during the preparation of this manuscript. It was used to suggest improvements to language and clarity, to propose \LaTeX{} edits, and to cross-check numerical values quoted in the text against the tables and figures. The authors reviewed every AI-suggested change and revised it where needed. All scientific content, analysis, and conclusions are the authors' own, and the authors take full responsibility for the final manuscript.

\end{acknowledgments}

\appendix
\section{Galactic Chemical Evolution}
\label{sec: GCE}

\subsection{Fe Production}
\label{sub: iron production}

    We consider two primary channels for iron production: core-collapse supernovae (CCSNe) and Type Ia supernovae (SNe~Ia).

    The CCSN channel is associated with the deaths of massive stars and is assumed to trace the star formation rate, $\psi(t)$ \citep{Madau_2017}, directly, reflecting the short lifetimes of massive star progenitors {($\sim 10\,\mr{Myr}$)} compared to Galactic evolution timescales. We adopt a fiducial iron yield of $m_{\rm Fe}=0.074^{+0.008}_{-0.008}\,M_\odot$ per event \citep{Maoz_2017} and normalize the local volumetric rate to $7.05\times10^4\,\mr{Gpc^{-3}\,yr^{-1}}$ \citep{Li_2011}.

    In contrast, SNe~Ia are delayed relative to star formation, so the rates are modeled through a convolution of a DTD and SFH. We assume an iron yield of $m_{\rm Fe}=0.7^{+0.44}_{-0.59}\,M_\odot$ per event \citep{regos2002edgelittypeiasupernovae, Maoz_2017}, a power-law DTD with slope $\alpha=1.0\pm 0.1$, and a minimum delay time $t_{\rm min}=400\,\mr{Myr}$ (typical time-scale for white-dwarf formation). The uncertainty of the minimum delay time for the SNe~Ia is large, spanning $40\,\mathrm{Myr}\sim1\,\mathrm{Gyr}$, but we have investigated various values and find no change in the conclusions.
    The local volumetric rate is assumed to be $2.55\times10^4\,\mr{Gpc^{-3}\,yr^{-1}}$ \citep{desai2026supernovaratesluminosityfunctions}. There are uncertainties in the ejecta of both SNe~Ia and CCSNe, but they are negligible compared to the uncertainties of BNS mergers. It is shown in \cite{Maoz_2017} that these assumptions retrieve the cosmic iron abundance history.
    Together, CCSNe and SNe~Ia determine the evolution of iron enrichment in the ISM.

\subsection{One-Zone Chemical Evolution Model}

    We model the chemical evolution of the interstellar medium (ISM) using the one-zone framework described in the Methods section \citep{Siegel_2019_1}. In this approach, the ISM is treated as a single well-mixed reservoir, such that newly produced elements are instantaneously and homogeneously distributed. This approximation neglects spatial inhomogeneities and stochastic enrichment at low metallicity, but captures the mean abundance evolution at intermediate and high metallicities. This is an appropriate approach since the metallicity of the disk of the Milky Way is observed to be well mixed from $\mr{[Fe/H]}\approx -1$ to $0.5$ \citep{Xiang_2022, cerqui2025chemicalenrichmenthistoriesmilky}.

The evolution of iron and europium masses are given by
\begin{equation}
\begin{split}
\frac{\mr{d}M_{\rm Fe}}{\mr{d}t} &= \overline{m}_{\mr Fe, \rm CCSN} R_{\rm CCSN}(t) \\
&\quad + \overline{m}_{\rm Fe, SN\,Ia} R_{\rm SN\,Ia}(t) - M_{\rm Fe} f(t)
\end{split}
\end{equation}
and
\begin{equation}
\begin{split}
\frac{\mr{d}M_{\rm Eu}}{\mr{d}t} &= \overline{m}_{\mr{Eu}, Delayed} f_{\rm NS} R_{Delayed}(t) \\
&\quad + \overline{m}_{\mr{Eu}, FM} f_{\rm NS} R_{FM}(t) - M_{\rm Eu} f(t).
\end{split}
\end{equation}

Here, $R_{Delayed}(t)$ and $R_{FM}(t)$ denote the merger rates of delayed and fast-merging BNSs, respectively. These rates are computed as convolutions of the star formation history with the corresponding delay time distributions (see Methods).

The parameter $f_{\rm NS}$ accounts for uncertainties in the fraction of BNS mergers contributing to \textit{r}-process enrichment and in the normalization of the local merger rate. We adopt a fiducial value $f_{\rm NS}=0.5$. The term $f(t)$ represents depletion of metals due to star formation and galactic outflows, which is elaborated in ~\cite{Hotokezaka_2018}.

\section{Bayesian Inference Results}

\subsection{Prior}
    The prior distributions of each parameter used in the inference are given in Table~\ref{tab:priors}.
    We use the GWTC-4.0 Full-Pop mass distribution from \cite{theligoscientificcollaboration2025gwtc40populationpropertiesmerging}, and estimate the average ejecta mass following \cite{Kr_ger_2020}. The short GRB DTD parameter posterior distributions provided by \cite{Zevin_2022} are used as the priors for the DTD parameters. The priors of enrichment efficiency parameters (Eq.~\ref{eq:efficiency}) are chosen to be uniform.
    
    \begin{deluxetable*}{lcc}
    \tablecaption{Prior distributions for the inference parameters.\label{tab:priors}}
    \tablehead{
    \colhead{Parameter} & \colhead{Description} & \colhead{Prior}
    }
    \startdata
     Local Rate   & Local Rate of Events  &  GWTC-4.0 Full-Pop \\
    $ \overline{m}_\mr{ej,0} $ & Average mass-ejecta of Event  &   Following \cite{Kr_ger_2020}  \\
    $ \alpha $   & Index of sGRB DTD power-law   & Posterior samples from \cite{Zevin_2022} \\
    $ t_\mr{min}$   & Minimum delay time of sGRB & Posterior samples from \cite{Zevin_2022} \\
    $\beta$   & Broken power-law low-$z$ index  & $U(-10, 10)$ \\
    $\delta $   & Broken power-law high-$z$ index  & $U(-10, 10)$ \\
    $z^\star$   & Broken power-law Cutoff & $U(0.5, 10)$ \\
    $f_{FM}$    & Local Rate Fraction of \textit{FM} Population & $U(0,1)$ \\
    \enddata
    \tablecomments{$U(a,b)$ denotes a uniform prior between $a$ and $b$.}
    \end{deluxetable*}

\subsection{Evidence}
\label{subsec: Evidence}

    \begin{deluxetable*}{lcccc}
    \tablecaption{The log-evidence, uncertainty of the log-evidence, and number of parameters for each scenario.\label{tab:log-evidence}}
    \tablehead{
    \colhead{Scenario} & \colhead{$\ln Z$} & \colhead{$\ln Z_{\rm err}$} & \colhead{Parameters} }
    \startdata
    $\mathcal{M}_1$ & $253.7$ & $0.16$   & $4$ \\
    $\mathcal{M}_2$ & $346.1$ & $0.16$   & $7$ \\
    $\mathcal{M}_3$ & $298.2$ & $0.11$   & $2$  \\
    $\mathcal{M}_4$ & $348.9$ & $0.16$   & $5$  \\
    $\mathcal{M}_5$ & $289.7$ & $0.15$   & $5$ \\
    $\mathcal{M}_6$ & $344.3$ & $0.18$   & $8$ \\
    $\mathcal{M}_7$ & $346.2$ & $0.16$   & $8$  \\
    $\mathcal{M}_8$ & $346.2$ & $0.17$   & $11$  \\
    $\mathcal{M}_9$& $345.3$ & $0.17$    & $8$  \\
    \enddata
    \label{tab:log-evidence}
    \end{deluxetable*}
We compute the Bayesian evidence using nested sampling \citep{skilling}.
The Bayesian evidence for scenario $\mathcal{M}_i$ is defined as
    \begin{equation}
    \mathcal{Z}_i \equiv \int \mathcal{L}(d \mid \vec{\theta}, \mathcal{M}_i)\,\pi(\vec{\theta} \mid \mathcal{M}_i)\, d\vec{\theta}.
    \end{equation}
    The definition of the Bayes factor to compare models $\mathcal{M}_i$ and $\mathcal{M}_j$ is
    \begin{equation}
    B^i_j \equiv \frac{\mathcal{Z}_i}{\mathcal{Z}_j}.
    \end{equation}
    
    Here, $\mathcal{L}(d \mid \vec{\theta}, \mathcal{M}_i)$ is the likelihood and $\pi(\vec{\theta} \mid \mathcal{M}_i)$ is the prior on the model parameters $\vec{\theta}$. Values of $B^i_j > 1$ indicate support for $\mathcal{M}_i$ over $\mathcal{M}_j$.

 Table~\ref{tab:log-evidence} shows the resulting log-evidence and the number of parameters for each model.

\section{Stochastic Gravitational-Wave Background Calculation}
\label{app:sgwb}

For each merger-rate evolution scenario, we compute the present-day
stochastic gravitational-wave background (SGWB) energy-density spectrum,
\begin{equation}
\Omega_{\rm GW}(f)
=
\frac{1}{\rho_{c,0}}
\frac{d\rho_{\rm GW}}{d\ln f},
\label{eq:sgwb-definition}
\end{equation}
where $f$ is the detector-frame frequency and
\begin{equation}
\rho_{c,0}
=
\frac{3H_0^2 c^2}{8\pi G}
\label{eq:critical-density}
\end{equation}
is the critical energy density of the Universe today.

We evaluate the standard cosmological integral
\begin{equation}
\Omega_{\rm GW}(f)
=
\frac{f}{\rho_{c,0} H_0}
\int dz\,
\frac{R(z)}{(1+z)E(z)}
\left.
\frac{dE_{\rm GW}}{df_s}
\right|_{f_s=(1+z)f},
\label{eq:sgwb-cosmological-integral}
\end{equation}
where $R(z)$ is the source-frame comoving BNS merger-rate density,
$f_s=(1+z)f$ is the source-frame frequency, and
\begin{equation}
E(z)
=
\sqrt{\Omega_m(1+z)^3+\Omega_\Lambda}.
\label{eq:cosmological-ez}
\end{equation}
We assume a flat $\Lambda$CDM cosmology throughout.

We model each source as an equal-mass $1.4\,M_\odot+1.4\,M_\odot$
binary and retain only the inspiral contribution to the emitted spectrum,
\begin{equation}
\frac{dE_{\rm GW}}{df_s}
=
\frac{\pi^{2/3}}{3}
G^{2/3}
\mathcal{M}_c^{5/3}
f_s^{-1/3},
\label{eq:inspiral-energy-spectrum}
\end{equation}
where the chirp mass is

\begin{equation}
\mathcal{M}_c
=
\frac{(m_1m_2)^{3/5}}{(m_1+m_2)^{1/5}}.
\label{eq:chirp-mass}
\end{equation}

With this approximation, the SGWB in the detector band follows the standard compact-binary inspiral power law,
\begin{equation}
\Omega_{\rm GW}(f)
=
\Omega_0
\left(\frac{f}{f_0}\right)^{2/3},
\label{eq:sgwb-power-law}
\end{equation}
where $f_0=25\,\mr{Hz}$ is the reference frequency used in this
work. The normalization $\Omega_0$ depends on the merger-rate evolution and therefore varies across the scenarios considered here. 
The cutoff frequency for the inspiral phase is determined by the Innermost Stable Circular Orbit (ISCO) \citep{Phinney:2001di}

$$f_\mr{ISCO} \approx \cfrac {c^3}{6^{3/2}\pi G M_\mr{tot}},$$

and converted to the detector frame for a given redshift.
We compute $\Omega_{\rm GW}(f)$ over the frequency range
$10$--$f_\mr{ISCO}/(1+z)\,\mr{Hz}$ and compare the resulting spectra with the latest LVK upper limits from the O1--O4a observing runs.

\section{Astrophysical Scatter}
    \label{sec:scatter}
    The likelihood evaluation in Equation~\eqref{likelihood}
    does not include intrinsic-scatter around the model track. It therefore propagates the observational uncertainties but does not account for astrophysical abundance variations unresolved by the one-zone model. To test the effect of this assumption, we repeated the inference after introducing an intrinsic-scatter parameter
    $\sigma_{\mr{int}}$
    in
    $\mr{[Eu/Fe]}$.
    
    For a model track
    $y=\mu_{\boldsymbol{\theta}}(x)$,
    where
    $x\equiv\mr{[Fe/H]}$
    and
    $y\equiv\mr{[Eu/Fe]}$,
    we assume that the true abundance is distributed around the mean track as
    \begin{equation}
        y_{\mr{true}}
        =
        \mu_{\boldsymbol{\theta}}(x)
        +\varepsilon_{\mr{int}},
        \qquad
        \varepsilon_{\mr{int}}
        \sim
        \mathcal{N}\left(0,\sigma_{\mr{int}}^2\right).
    \end{equation}
    
    The intrinsic-scatter is treated as an unobserved variable and marginalized over in the likelihood. For the observed abundance
    $\mathbf{d}_i=(x_i,y_i)^{\mr{T}}$
    of the
    $i$th star, the modified likelihood is
    \begin{align}
    \mathcal{L}_i(\boldsymbol{\theta}, \sigma_{\mr{int}})
    &= \int \mr{d}x \int \mr{d}\varepsilon_{\mr{int}}\,
    \mathcal{N}\!\left(
        \mathbf{d}_i \,\middle|\,
        \big(x,\ \mu_{\boldsymbol{\theta}}(x) + \varepsilon_{\mr{int}}\big)^{\!\top},\,
        \boldsymbol{\Sigma}_i
    \right) \nonumber \\
    &\quad\times
    \mathcal{N}\!\left(
        \varepsilon_{\mr{int}} \,\middle|\, 0,\ \sigma_{\mr{int}}^2
    \right)
    \nonumber \\[4pt]
    &= \int \mr{d}x\,
    \mathcal{N}\!\left(
        \mathbf{d}_i \,\middle|\,
        \big(x,\ \mu_{\boldsymbol{\theta}}(x)\big)^{\!\top},\,
        \boldsymbol{\Sigma}_i^{\mr{eff}}
    \right).
    \label{eq:scatter_likelihood}
\end{align}
    Here, the effective covariance matrix is
    \begin{equation}
    \boldsymbol{\Sigma}_i^{\mr{eff}}
    =
    \boldsymbol{\Sigma}_i+
    \begin{pmatrix}
    0 & 0\\
    0 & \sigma_{\mr{int}}^2
    \end{pmatrix}
    =
    \begin{pmatrix}
    \sigma_{x,i}^2 & 0\\
    0 & \sigma_{y,i}^2+\sigma_{\mr{int}}^2
    \end{pmatrix}.
    \label{eq:scatter_covariance}
    \end{equation}
    
    Thus, marginalizing over the intrinsic-scatter is equivalent to adding
    $\sigma_{\mr{int}}^2$
    to the observational variance in
    $\mr{[Eu/Fe]}$,
    while leaving the uncertainty in
    $\mr{[Fe/H]}$
    unchanged. We treat
    $\sigma_{\mr{int}}$
    as an additional parameter and infer it jointly with the GCE model parameters.
    
    \subsection{Results Adding Intrinsic-Scatter Term}
    Including the intrinsic-scatter term, the log-evidence for the evolving scenarios has decreased, narrowing the gap between them (Table~\ref{tab:log_evidence_scatter}). Comparing $\mathcal{M}_1$ and $\mathcal{M}_2$, this corresponds to a change in the log-Bayes factor from $B^{2}_{1} > 10^{40}$ without the intrinsic-scatter term to $B^{2}_{1} > 10^{37}$ with the scatter term included. This is consistent with the role of $\sigma_{\rm int}$: because the original likelihood (Eq.~\ref{likelihood}) attributes all deviation from the model track to observational uncertainty, tracks that pass closer to the centroid of the data are favored more strongly than the astrophysical scatter in the data would otherwise justify. Despite this reduction, the preference for evolving over non-evolving scenarios remains decisive. We conclude that the qualitative preference for evolving scenarios reported in the main text is robust to this more conservative treatment of the likelihood.
    \begin{deluxetable*}{lc}
    \tablecaption{The log-evidence for each scenario with an intrinsic-scatter parameter.}
    \label{tab:log_evidence_scatter}
    \tablehead{
    \colhead{Scenario} &
    \colhead{$\ln Z$ (with intrinsic-scatter)}
    }
    \startdata
    $\mathcal{M}_1$ & $254.6$ \\
    $\mathcal{M}_2$ & $341.7$ \\
    $\mathcal{M}_3$ & $296.3$ \\
    $\mathcal{M}_4$ & $345.0$ \\
    $\mathcal{M}_5$ & $288.0$ \\
    $\mathcal{M}_6$ & $340.8$ \\
    $\mathcal{M}_7$ & $342.0$ \\
    $\mathcal{M}_8$ & $341.7$ \\
    $\mathcal{M}_9$ & $341.4$ \\
    \enddata
    \end{deluxetable*}

\bibliography{bib}
\bibliographystyle{aasjournalv7}

\end{document}